\documentclass[%
 reprint,
 amsmath,amssymb,
 aps,
]{revtex4-2}

\usepackage{graphicx}
\usepackage{dcolumn}
\usepackage{bm}
\usepackage{physics}
\usepackage{hyperref}
\usepackage{color}

\newcommand{\beq}{\begin{equation}}
 \newcommand{\eeq}{\end{equation}}
 \newcommand{\bel}{\begin{align*}}
 \newcommand{\tamam}{\end{align*}}

 \newcommand{\beqa}{\begin{eqnarray}}             
 \newcommand{\eeqa}{\end{eqnarray}}               
 \newcommand{\nn}{\nonumber}                      

\begin{document}

\title{Violating Bell inequality using weak coherent states}

\author{Moslem Mahdavifar}
\email{mahdavifar.m@miami.edu}
\affiliation{Department of Physics, University of Miami, Coral Gables, Florida 33146, USA}
\author{S. M. Hashemi Rafsanjani}

\affiliation{Department of Physics, University of Miami, Coral Gables, Florida 33146, USA}

\renewcommand{\figurename}{Fig.}

\def\pmid{p_{mid}}
\def\SIn{\sigma_\pm^{in}}
\def\SOut{\sigma_\pm^{out}}
\def\SD{D_{\pm}}
\def\SH{h_{\pm}(x,t)}

\begin{abstract}
We present an experimental investigation of two-photon interference using a continuous-wave laser. We demonstrate the violation of the CHSH inequality using the phase randomized weak coherent states from a continuous wave laser. Our implementation serves as an approach to reveal the quantum nature of a source that is considered to be a \textit{classical} source.
\end{abstract}

\maketitle

\newpage

\section{Introduction}
Non-classical states of light 
 frequently appear in quantum-enabled applications such as quantum metrology, quantum teleportation, quantum computation, and metrology \cite{bouwmeester1997experimental, knill2001scheme, giovannetti2011advances, RevModPhys.84.777, pirandola2015advances}.
The conventional approach to realizing such states is to employ nonlinear optics. However, because optical nonlinearities have small couplings, usually, the production rate of resultant photons is orders of magnitude smaller than the rate of photons in the pump laser \cite{boyd2020nonlinear}. 
This limited rate is an acute limitation because the application of Non-classical states in the forementioned applications typically requires multi-photon interference, which is extremely sensitive to photon loss.
As the number of participating photons increases, this problem grows rapidly and quickly renders the experiment infeasible \cite{bachor2004guide}. It is thus no surprise that the genuine interference of mere three photons was directly observed only recently \cite{PhysRevLett.118.153602, PhysRevLett.118.153603}.

Recently, several studies have suggested that phase-randomized weak coherent states (PRWCS) are an alternative candidate for single photons in many experiments \cite{PhysRevA.94.062305, Navarrete_2018}. For instance, Valente and Lezama implemented quantum state tomography of phase-randomized coherent states and observed the Wigner function negativity of a single photon \cite{Valente:17}. These states have also appeared in measurement device-independent quantum key distribution \cite{Islam_2019}. Furthermore, the quintessential example of two-photon interference, i.e., the Hong-Ou-Mandel effect, has been observed using phase randomized coherent states too \cite{Aragoneses:18, Arruda:20}. One immediate benefit of using PRWCS is to avoid the limitation on the number of photons due to the low efficiency of optical non-linearities. However, in this approach, the signal from possible competing effects is added to the relevant signal. One needs to repeat the experiment with different inputs to isolate the relevant signal and remove the contribution of undesirable states. 

There is no fundamental reason that the promise of the phase-randomized coherent states should be limited to single-photon states or studies that only involve measuring one interferometric signal. One would expect that phase-randomized single-photons can be engineered to create non-classical states, and the contribution of these non-classical states in a linear optics experiment could also be isolated and measured. In this work, we present such an example by demonstrating the violation of the Bell inequalities using the phase-randomized weak coherent states of light. 

The introduction of Bell inequality made it possible to discern, experimentally, between the quantum mechanical description and realist hidden variable hypotheses of nature \cite{bell2004speakable, PhysRevLett.23.880}. Following a series of successful early experiments \cite{PhysRevLett.49.1804, PhysRevLett.47.460}, the Bell inequalities have been violated using various physical systems including photons pairs\cite{pan2000experimental}, entangled diamond spins \cite{hensen2015loophole}, Josephson phase qubits \cite{ansmann2009violation}, and under an increasingly stringent assumptions \cite{rowe2001experimental, PhysRevLett.100.220404, giustina2013bell, PhysRevA.90.032107, PhysRevLett.118.060401, PhysRevLett.119.010402, PhysRevLett.121.080403}. While previous experiments used a source that primarily produces entangled photons, we measure the entangled state's contribution by subtracting the unwanted contribution of irrelevant separable states. Violating Bell inequality is also a direct way to certify the entanglement between separate photons, which could subsequently be utilized for other applications.

In this work, we report on violation of the CHSH inequality \cite{PhysRevLett.23.880} using phase-randomized weak coherent states of a continuous-wave laser. To achieve this, we create Maximally entangled Bell states by manipulating the phase-randomized coherent states from a continuous-wave laser. We then implemented the correlated local measurements on the two separate photons of this Bell state and measured the Bell state's contribution from its mixture with other undesirable states. In the following, we will first describe the theory behind our experiment, how to create the Bell state from phase-randomized coherent states, the byproduct created states of this process, and how to isolate and measure the contribution of the Bell state to the rate of two-photon events that constitute the correlated local measurement that we need to observe the violation of the Clauser-Horne-Shimony-Holt (CHSH) formulation of the Bell inequality \cite{PhysRevLett.23.880, PhysRevLett.47.460, PhysRevLett.49.91}. We will then present our experimental results and finish with a brief discussion of the implications of our work for prospective studies. 

\section{Bell state and phase-randomized weak coherent states of light}

We demonstrate the violation of the CHSH inequality for a Bell state. The Bell states are a set of four two-qubit maximally-entangled states \cite{nielsen2002quantum}. The photon-based Bell states are usually produced by employing optical non-linearities. However, these states also appear in linear optics. In the following, we explain how we use phase-randomized weak coherent states to create a Bell state and isolate its contribution from an incoherent mixture with other quantum states. 

To produce a Bell state, we will use a beam splitter to split the beam from a coherent light source into two beams. We will then impress a random phase onto one of the split beams and rotate the polarization of the other beam into an orthogonal polarization. Next, we recombine the two beams at a secondary beam splitter. The quantum state that describes the two beams before their recombination is 
\begin{align}
\rho = \sum_{i,j} P_{a}(i)P_{b}(j) \ket{i_{aH},j_{bV}}\bra{i_{aH},j_{bV}}, \label{phase-random}
\end{align}
where the probability distributions $P_a(i), P_b(j)$ are Poisson distributions:
\begin{align}
P_s(n) = \frac{\mu_s^n}{n!}e^{-\mu_s}, \hspace{0.5 cm} s = a,b.
\end{align}
Here, $\mu_a, \mu_b$ are the mean photon number at each input port of the secondary beam splitter. The state $\ket{i_{aH},j_{bV}}$ denotes a state with $i$ photons of polarization $H$ at the input port $a$ and $j$ photons of polarization $V$ at the input port $b$. 
To see why equation (\ref{phase-random}) describes the state accurately, notice that in a continuous wave laser the phase of the coherent state changes a between the different temporal modes. Thus over many temporal modes, the quantum state of each beam before recombination is given by a mixed state because \cite{sargent1974laser, PhysRevLett.88.027902}:
\begin{align}
\int \frac{d\phi}{2\pi} \ket{\sqrt{\mu} e^{i\phi}}\bra{\sqrt{\mu} e^{i\phi}} =  \sum_{n=0}P_n(\mu) \ket{n}\bra{n}.
\end{align}
In addition, the role of the explicit phase-randomization that we introduce is to make the Poisson distributions of the two recombining beams independent.
In the weak light limit, i.e. $\mu_s \ll 1$, the contribution from higher photon number  states is negligible compared to the contribution from the two-photon states and we will ignore those contributions in our following analysis. As our measurements are all two-photon events, we can safely ignore the single-photon contribution to the quantum state. The two-photon component of the above mixed state is 
\begin{align}
\ket{\Psi_2}&\propto  \mu_a\mu_b \ket{1_{aH},1_{bV}}\bra{1_{aH},1_{bV}}   \\ \nn
&+\frac{\mu_a^2}{2!} \ket{2_{aH},0_{bV}}\bra{2_{aH},0_{bV}} 
+\frac{\mu_b^2}{2!} \ket{0_{aH},2_{bV}}\bra{0_{aH},2_{bV}} \label{equation:state}
\end{align}

If we denote the output ports of the beam splitter as $c$ and $d$, the three input beams are converted to the following states at the output.
\begin{align}
&\ket{1_{aH},1_{bV}} \rightarrow \frac{1}{\sqrt{2}}\left(\ket{\psi_-}+ i\ket{\phi_+}\right),\\ \nn
&\ket{2_{aH},0_{bV}} \rightarrow \frac{1}{2}\left(\ket{2_{cH},0_{dH}}-\ket{0_{cH},2_{dH}}+2i\ket{1_{cH},1_{dH}}\right),\\ \nn
&\ket{0_{aH},2_{bV}} \rightarrow \frac{1}{2}\left(\ket{0_{cV},2_{dV}}-\ket{0_{cV},2_{dV}}+2i\ket{1_{cV},1_{dV}}\right).
\end{align}
Here $\ket{\psi_-}$ and $\ket{\phi_+}$ are two of the Bell states:
\begin{align}
\ket{\psi_{\pm}} = \frac{1}{\sqrt{2}}\bigl(  \ket{1_{cH}, 1_{dV}} \pm \ket{1_{cV}, 1_{dH}} \bigr),  \\
\ket{\phi_{\pm}} = \frac{1}{\sqrt{2}}\bigl(  \ket{1_{cH}, 1_{cV}} \pm \ket{1_{dH}, 1_{dV}} \bigr).
\end{align}

One important point is that among the above states only $\ket{\psi_{-}}$, $ \ket{1_{cH}, 1_{dH}} $, and 
$\ket{1_{cV},1_{dV}}$ leads to two-photon detection events with detection of one photon at a separate output of the beam splitter. The rest of the states only cause two-photon events at the same output of the secondary beam splitter. This is important because all the required correlations for the CHSH inequality are of the former kind, and the latter kind does not contribute to any of our measurements. Therefore only the three states $\ket{\psi_{-}}$, $ \ket{1_{cH}, 1_{dH}} $, and $\ket{1_{cV},1_{dV}}$ do contribute any of our correlations measurements, and our immediate problem is how to isolate the contribution from the $\ket{\psi_{-}}$ from its mixture with the remaining separable states. Nevertheless, because they are incoherently mixed with our relevant states ($\ket{\psi_{\pm}}$), we can measure their contribution separately and subtract from the signal.

The process for the isolation and measurement of the relevant signal in a mixed states follows closely the technique that was originally proposed in \cite{PhysRevA.94.062305} and was applied recently in \cite{Aragoneses:18, Arruda:20} to demonstrate two-photon interference. 
The CHSH inequality is 
\begin{align}
{\mathcal{S} =} |E(\alpha, \beta)-E(\alpha, \beta')+E(\alpha', \beta)+E(\alpha', \beta')|\le 2,
\end{align}
where $\alpha,\alpha'$ linear polarization rotation angles at detector A, and $\beta,\beta'$ are the linear polarization rotation angles at detector B. 
To measure the CHSH inequality one needs to quantify the normalized quantum correlations:
\begin{align}
E(\alpha, \beta) = \frac{C_{++}-C_{+-}-C_{-+}+C_{--}}{C_{++}+C_{+-}+C_{-+}+C_{--}}.
\end{align}
where $C_{i,j}$'s are the rate of two-photon events with polarizations $i,j$ at the two detectors respectively. $(\alpha,\beta)$ are the linear rotation angles for polarization bases at each detector. For the singlet state the normalized quantum correlation function only depends on the difference between the difference between the two angles $E(\alpha, \beta) = E(\alpha- \beta)$. 

In addition, each $C_{i,j}$ can be quantified using three different measurements. To see this notice that
\begin{align}
C_{i,j} \propto P_{i,j}(1_d,1_c|1_a,1_b),
\end{align}
where $P_{i,j}(1_d,1_c|1_a,1_b)$ is the probability of having one photon leaves from each output of the beam splitter conditioned upon one photon enters each input port. The same probability also appears in the total rate of two-photon events with two photons with polarizations $i,j$ at respective detectors:
\begin{align}\nn 
&\mathcal{N}_{i,j}(\mu_a,\mu_b) =  \mu_a\mu_b P_{i,j}(1_d,1_c|1_a,1_b) +\\ 
&\frac{\mu_a^2}{2}  P_{i,j}(1_d,1_c|1_a,0_b)  + \frac{\mu_b^2}{2} P_{i,j}(1_d,1_c|0_a,1_b).
\end{align}
The two extra conditional probabilities also appear in the same equation if one of the input photons is blocked: 
\begin{align}\nn 
\mathcal{N}_{i,j}(\mu_a,0) =  \frac{\mu_a^2}{2}  P_{i,j}(1_d,1_c|2_a,0_b), \\ 
\mathcal{N}_{i,j}(0,\mu_b) =  \frac{\mu_b^2}{2}  P_{i,j}(1_d,1_c|0_a,2_b).
\end{align}
Thus by subtracting these two contributions from the original coincidence term, we can isolate the contribution from the Bell state:
\begin{align}
C_{i,j} = \mu_a\mu_b P_{i,j}&(1_d,1_c|1_a,1_b) \\ \nn \simeq &\mathcal{N}_{i,j}(\mu_a,\mu_b) - \mathcal{N}_{i,j}(\mu_a,0)  - \mathcal{N}_{i,j}(0,\mu_b) \label{eqn_norm}
\end{align}
One benefit of this approach is that the computed probabilities are normalized by construct, and one does not need to calibrate the conditional probabilities $P_{i,j}$'s directly.
Having devised an approach to measure correlations $E(\alpha, \beta)$ between any two angles, we can compute the contributing terms for CHSH inequality and observe the violation of this inequality for a maximally entangled state directly. 

\begin{figure}[t!]
    \centering {\includegraphics[width = 0.95\columnwidth]{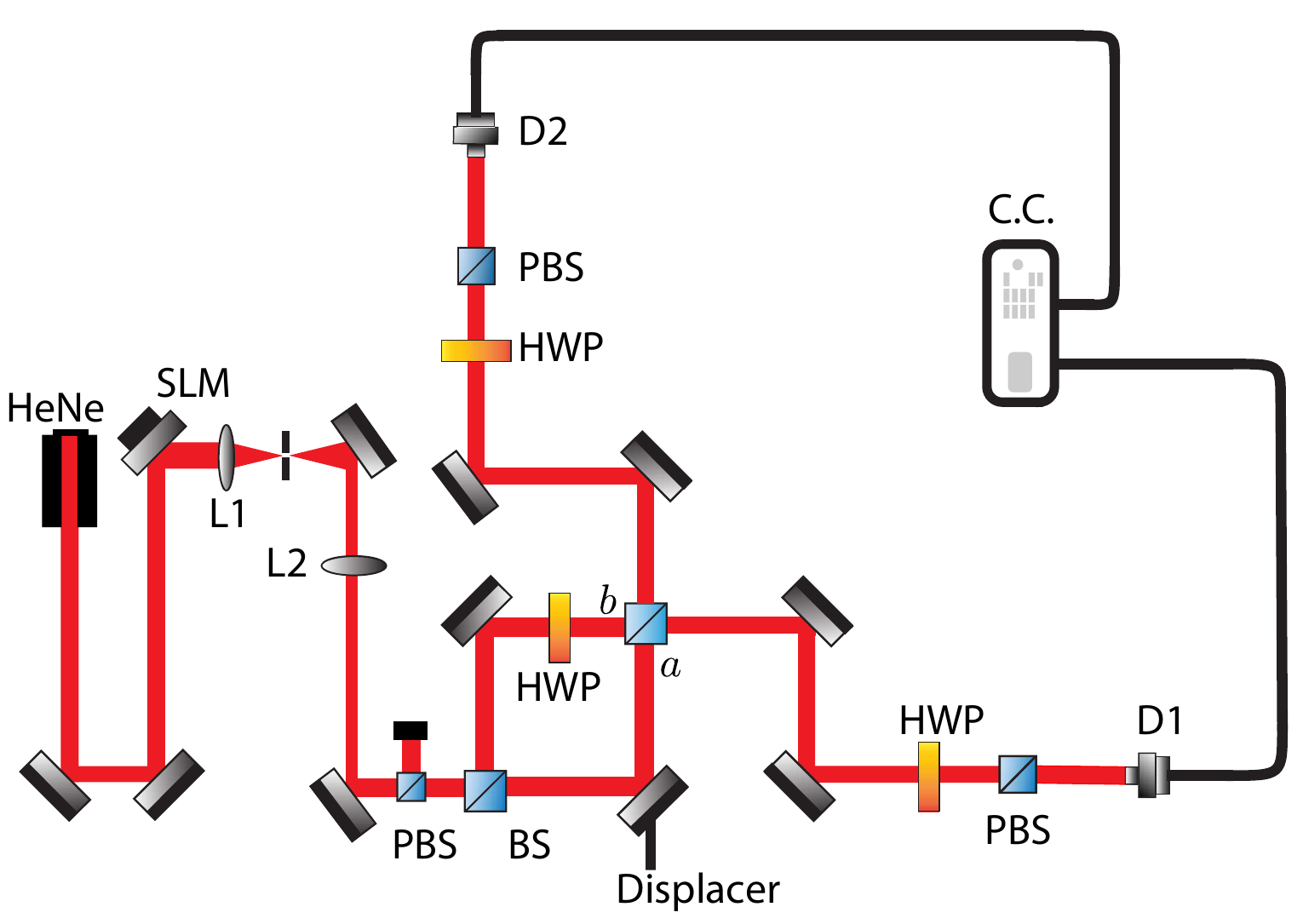}}
    \caption{A schematic representation of the experimental setup: Light from a HeNe laser is collimated and passed through a polarizing beam splitter (PBS). Then we use a balanced beam splitter to split the beam in two, and phase-randomization is added in the lower arm, while a half-wave plate (HWP) rotates by $90^\circ$ the polarization in the upper arm that is fed into port $b$ os the secondary beam splitter. The polarization in each output port is measured by subsequent combinations of HWP and polarizing beam splitter.}
    \label{FIGURE_setup}
\end{figure}

\section{Experiment and results}
Having explained the theory behind our technique, we next discuss our experiment and present our result. We use a continuous wave HeNe laser as the light source for this experiment. Fig. \ref{FIGURE_setup} shows a schematic representation of the experimental setup. The light from the laser is fed into a single-mode fiber to ensure the spatial single-mode nature of the beam. By controlling this coupling, we can further control the average photon number in the experiment. In successive order, we use a lens to collimate the beam that emerges from the other end of the fiber, employ a spatial light modulator to flatten the wavefront, and pass it through a polarizer to ensure its uniform polarization. A 50:50 balanced beam splitter then splits the beam in two, and we use a half-wave plate (HWP) to rotate the polarization of one of the resultant beams by $90^\circ$. A linear displacement ramp induces a slight random delay into the path of the second split beam. The delay induces a phase shift of up to $100\pi$. Nevertheless, the displacement is much smaller than the coherence length of the source to preserve the indistinguishability of the photons of the split beams. Thus the two-mode input quantum state to the secondary beam splitter is given by Eq.\,(\ref{equation:state}). 

We then recombine the two beams at a secondary beam splitter. Each of the secondary beam splitter outputs is passed through a variable polarization measurement where we use a HWP and a polarizing beam splitter to project onto either up or down linear polarization at specific angles. Because we are interested in violation of the CHSH inequality, all of our measurements are two-photon coincidence events between the two output ports of this secondary beam splitter. As we mentioned, the portion of the resultant states consisting of two-photon events in one of the output beams does not register in our measurements and will ignore them altogether. The restricted quantum state is a mixed state of three pure states: a singlet Maximally entangled state, and two separable states whose contribution we can identify and remove in our analysis.

\begin{figure}[bt!]
    \centering 
    \includegraphics[width = 0.95\columnwidth]{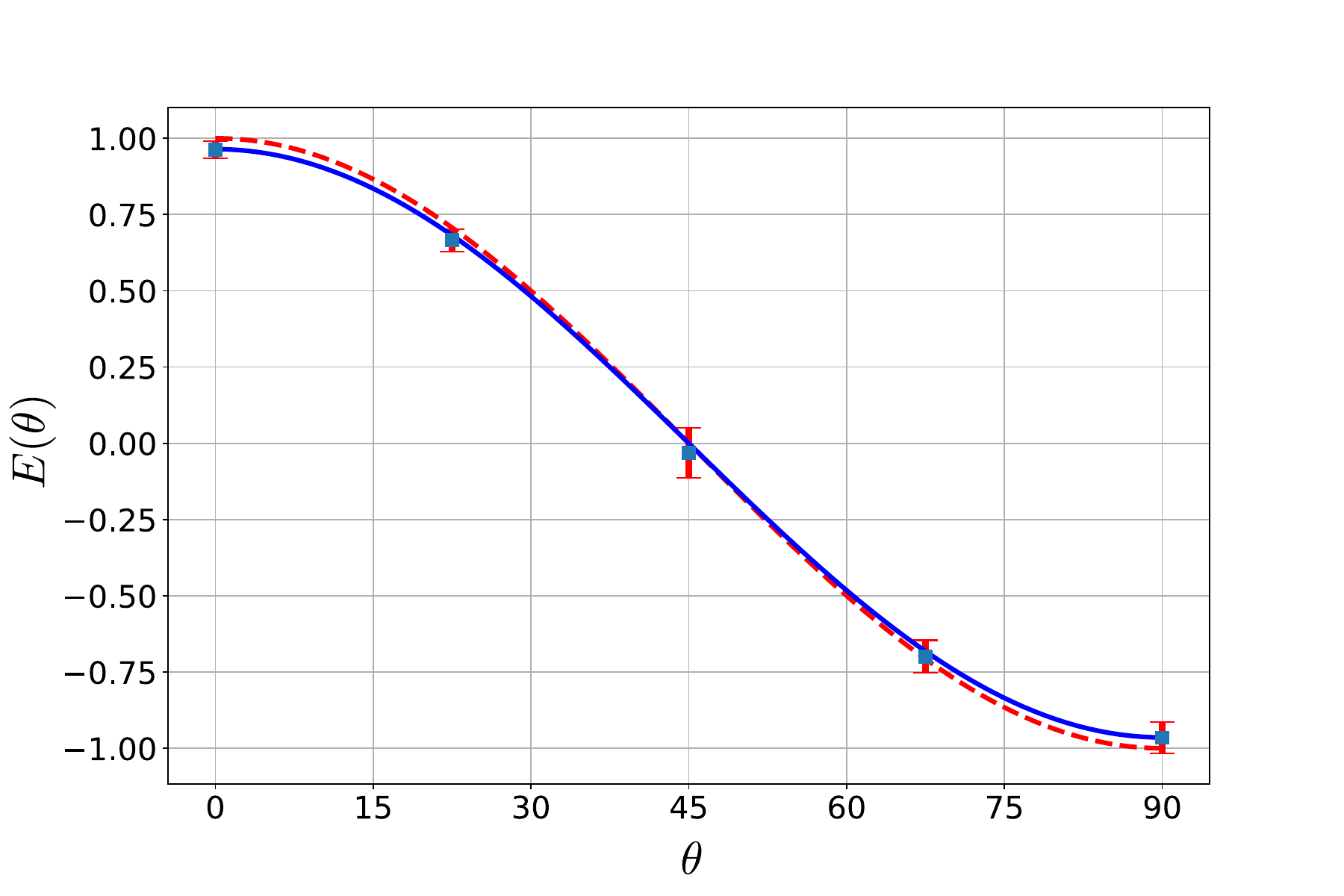}
    \caption{ The correlation function $E(\theta)$ as a function of the difference angle between the two polarization projections. The measured data is represented by blue dots with error bars representing one standard deviation. The dashed red line represents the ideal scenario and the solid blue line represented the best fit.}
    \label{FIGURE_data}
\end{figure}

We use the correlated measurements of  $\mathcal{N}_{i,j}(\alpha, \beta)$ to extract the correlation contribution $C_{i,j}$ to $E(\alpha,\beta)$ for the maximally entangled singlet state according to a recipe that we laid out before. For the singlet state $E(\alpha,\beta)=E(|\alpha-\beta|)$ and it only depends on the angular difference between the two states. The so-called \textit{Bell test angles}, for which the violation reaches its maximum value are  $(\alpha,\alpha',\beta,\beta') = (0,\frac{\pi}{4}, \frac{\pi}{8}, \frac{3\pi}{8})$.
For this choice of angles the CHSH inequality reads:
\begin{align}
\mathcal{S} = |3E(\theta)-E(3\theta)|\le 2, \quad \theta = \frac{\pi}{8}
\end{align} 
The result for the measurement of the correlation function $E(\theta)$ is presented in Fig.\,\ref{FIGURE_data}. The dashed line shows the ideal theoretical prediction for a maximally entangled singlet state. Each of the blue dots are accompanied by a red bar representing one standard deviation of ten measurements. The blue solid line represents the best fit of the data to $E(\theta) = \eta \cos\theta$. An $\eta<1$ arises because of the decoherence of the entangled state. The best fit to our data is $\eta = 0.964$, which is consistent with the excellent initial agreement of the experimental with the result of an ideal singlet state.

The maximal violation of the CHSH inequality occurs at $\theta = \frac{\pi}{4}$. If we use the measured values of $E(\theta)$ from our measurement one can readily calculate the value of 
\begin{align}
\mathcal{S} = 2.70\pm0.13
\end{align}
Note that inclusion of the realistic decoherence also affects that maximum value of the violation to $\mathcal{S}_{max} = 2\sqrt{2}\eta = 2.72$. Our measurement constitutes a violation of the CHSH inequality of more than five standard deviation.

Two comments about our result are in order. First, we would like to make a distinction between our results here, and some of the recent works on experimental violations of Bell inequality using so-called \textit{classical light} \cite{kagalwala2013bell, qian2015shifting}. These previous studies use the non-separability between different degrees of freedom to mimic entanglement and its behavior. Violation of Bell's inequality in this context signals the non-separability of various degrees of freedoms of the beam, or even one photon. In contrast, the violation of the Bell inequality in our work is the quantum entanglement between two remote photons that share a maximally entangled singlet state. This is the same quantum state that was originally exploited by Aspect et al., 
in their seminal work to show the violation of the Bell inequalities \cite{PhysRevLett.49.1804}. 

The distinction between our approach and past efforts is in the production of the entangled state. Conventional schemes engineer a source that primarily produces maximally entangled states \cite{PhysRevLett.49.1804, pan2000experimental, rowe2001experimental, PhysRevLett.100.220404, giustina2013bell, PhysRevA.90.032107, PhysRevLett.118.060401, PhysRevLett.119.010402, PhysRevLett.121.080403}. Such sources guarantee that the measurement results are dominated by the Bell state contribution. Producing correlated photons, however, becomes exceedingly more challenging as the number of photons in correlation increases. In contrast, we devised a strategy to isolate and and measure the contribution of a Bell state from its mixture with other states. This strategy then allows us to forgo the need for a pure maximally entangled source. Thus one no longer needs a source of pure correlated photons, but can employ a source that produces a mixture of the entangled state with other states. We are thus able to use a phase-randomized coherent states from a laser directly, which provides many orders of magnitudes more initial photons for the experiment. Extracting the contribution of a specific state from a mixture requires repeated measurements with different input to isolate remove the contribution form undesirable states, but, as evidenced by our work, this isolation and extraction can be successfully undertaken across different measurements and the combined measured quantity still corresponds to the expected result for a pure maximally entangled state.

Secondly, We focused our effort on isolating and measuring the contribution of a Bell state of two remote photons from data that include the contribution from other quantum states. We did not spend any effort here to make sure that the remote measurements were space-like or randomly chosen. Thus our experiment does not constitute a loop-hole-free example of the violation of Bell's inequality. What it demonstrates instead is an experimental confirmation of the entanglement between two photons that were recombined at the secondary beam splitter, isolated, and observed out of its mixture with separable states. 

\section{Conclusion}
In this work, we demonstrated the first violation of CHSH inequality using phase-randomized coherent states. To achieve this, we produced a Bell state by combining two independently phase-randomized coherent states. The scheme also leads to the production of unwanted separable states whose contribution we successfully removed from the final result. 

Our demonstration opens the door to the possibility of using phase-randomized coherent states to produce Non-classical states of even higher photon numbers. As we created non-classical states and measured their contribution with phase-randomized coherent states, it is reasonable to expect that a similar approach can produce non-classical states of higher photon numbers. This opening then is an excellent opportunity for technologies that employ multi-photon states. e.g., metrology, quantum computation. One can imagine feeding phase-randomized photons into other optical setups to produce other non-classical states (e.g., NOON states) and then deploying these states for in an experiment (e.g., phase-measurement) \cite{giovannetti2011advances, PhysRevLett.85.2733, PhysRevA.65.052104, Krenn14197}. While our target non-classical state will be mixed with other possible undesirable states, we can deploy a similar approach to isolate and measure the contribution of the target state to the signal in the experiment. This possibility exists even if the desired signal, as in the case of Bell inequality violation, is composed of several distinct quantities.

The above point is singularly noteworthy for applying the transverse structure of light in quantum optical experiments \cite{HashemiRafsanjani:19}. While the spatial properties of photons present an excellent medium for realizing multi-photon interference, they have been chiefly shunned due to high loss in production and manipulation of the transverse profile of photons \cite{PhysRevA.45.8185, PhysRevLett.105.153601, PhysRevLett.116.130402}. With phase-randomized coherent states, one is no longer bound by the relatively small number of correlated photons that are sourced for the experiment, and the experiments can tolerate a much higher loss. In effect, photon-loss is not a prohibitive bar to render such studies infeasible. Finally, of practical importance is the simplicity of our system. The reliance on phase-randomized coherent states allowed us to forgo the need for a pulsed light source, often required in similar experiments. We successfully demonstrated violation of the CHSH inequality using a HeNe laser, which is more cost-effective than a typical alternative for similar experiments. We believe this is a significant advantage that opens the door to the availability of such studies to a broader community. \\


\section*{Conflict of Interests}
The authors declare no conflict of interests.

\section*{Data availability} Data underlying the results presented in this paper are not publicly available at this time but may be obtained from the authors upon reasonable request.



%


\end{document}